\documentstyle{article}
\begin{document}

\begin{center}
{\Large {\bf Doorway states in nuclear reactions as a manifestation
of the ``super-radiant" mechanism}}
\end{center}

\begin{center}

N. Auerbach$^{1}$ and V. Zelevinsky$^{2}$\\[0.5cm]
{\small {\sl $^{1}$School of Physics and Astronomy, Tel Aviv
University, Tel Aviv 69978, Israel}\\
{\sl $^{2}$ National Superconducting Cyclotron Laboratory and
Department of Physics and Astronomy, Michigan State University, East
Lansing, Michigan 48824-1321, USA}}\\[1cm]

\end{center}

{\small {\bf Abstract}. A mechanism is considered for generating
doorway states and intermediate structure in low-energy nuclear
reactions as a result of collectivization of widths of unstable
intrinsic states coupled to common decay channels. At the limit of
strong continuum coupling, the segregation of broad
(''super-radiating") and narrow (''trapped") states occurs revealing
the separation of direct and compound processes. We discuss the
conditions for the appearance of intermediate structure in this
process and doorways related to certain decay channels.}

\newpage

\section{Introduction}

Apart from the two extremes of nuclear many-body dynamics, very
narrow compound resonances and broad single-particle or giant
resonances, nuclear cross sections exhibit other structures in
between. The intermediate structures are related to more organized
dynamics of nuclei as compared to the chaotic motion at the compound
stage. There is a large variety of possible motion that gives rise
to a diversity of intermediate structures.

The intermediate structure resonances in the nuclear cross sections
have usually widths that are satisfying the following inequality:
\begin{equation}
\Gamma_{q}\ll\Gamma_{{\rm in}}<\Gamma_{{\rm s.p.}}. \label{1}
\end{equation}
Here $\Gamma_{q}$ denotes a width of a typical compound state,
$\Gamma_{{\rm s.p.}}$ that of a single-particle resonance, and
$\Gamma_{{\rm in}}$ of the intermediate structure resonances.

Intermediate structure was often studied in the past using various
approaches \cite{griffin66,feshbach80,friedman81,gales88,BBB98},
recently with the aid of the ideas of quantum chaos and wavelet
analysis \cite{kilgus87,shevchenko04,kalmykov06}. In the present
work we apply the {\sl super-radiant} (SR) approach
\cite{SZNPA89,VZWNMP04} to show how the intermediate structure
emerges naturally in the theory. The SR approach has been used in
many different fields. For example, it was applied to problems in
chemistry \cite{verevkin88,remacle90}, atomic physics \cite{FGG96},
condensed matter physics \cite{pichugin01,VZWNMP04}, intermediate
energy nuclear physics \cite{auerbach94,AZ02,AZV04}, and most widely
in the theory of nuclear reactions
\cite{SZNPA89,rotter91,izr94,VZWNMP04}, in particular to describe
the chaotic dynamics of the compound nucleus and phenomena on the
borderline of discrete and continuum spectra in loosely bound nuclei
\cite{VZcont,VZCSM05}. In the next section we will develop the SR
formalism for some simple situations in order to arrive in a
heuristic manner at the concept of intermediate structure.

\section{The SR mechanism and nuclear dynamics}

\subsection{The effective Hamiltonian}

Following the standard projection formalism \cite{feshbach5862}, let
us divide the Hilbert space of nuclear states into two parts, the
$\{Q\}$-subspace involving very complicated many-body states
$|q\rangle$, and the subspace $\{P\}$ of open channels $|c\rangle$.
We use the notations ${\cal Q}$ and ${\cal P}$ for the corresponding
projection operators onto the above subspaces. The total wave
function of the system,
\begin{equation}
|\Psi\rangle={\cal Q}|\Psi\rangle+{\cal P}|\Psi\rangle, \label{2}
\end{equation}
satisfies the stationary Schr\"{o}dinger equation
\begin{equation}
H|\Psi\rangle=E|\Psi\rangle,                         \label{3}
\end{equation}
that can be decomposed into a set of coupled equations,
\begin{equation}
(E-H_{QQ}){\cal Q}|\Psi\rangle=H_{QP}{\cal P}|\Psi\rangle, \label{4}
\end{equation}
and
\begin{equation}
(E-H_{PP}){\cal P}|\Psi\rangle=H_{PQ}{\cal Q}|\Psi\rangle, \label{5}
\end{equation}
where we use the notations $H_{PQ}={\cal P}H{\cal Q}$ and so on.
Eliminating the part ${\cal P}|\Psi\rangle$, we come to the equation
in the $Q$-space,
\begin{equation}
(E-{\cal H}_{Q}){\cal Q}|\Psi\rangle=0,               \label{6}
\end{equation}
with the effective Hamiltonian
\begin{equation}
{\cal H}_{Q}=H_{QQ}+H_{QP}\,\frac{1}{E^{(+)}-H_{PP}}\,H_{PQ}.
                                                   \label{7}
\end{equation}
Here $E^{(+)}\equiv E+i0$ contains the infinitesimal imaginary term
$+i0$ ensuring right asymptotic conditions for the continuum wave
functions.

The second term of the effective Hamiltonian (\ref{7}) contains a
real and imaginary part of the propagator
\begin{equation}
G^{(+)}(E)=\frac{1}{E^{(+)}-H_{PP}}               \label{8}
\end{equation}
emerging from the principal value (off-shell contributions) and the
delta-function $\delta(E-H_{PP})$ (on-shell contributions from
channels $c$ open at energy $E$), respectively. The imaginary part,
$-(i/2)W$, of the effective Hamiltonian is given by
\begin{equation}
W=2\pi\sum_{c;{\rm open}}H_{QP}|c\rangle\langle c|H_{PQ}. \label{9}
\end{equation}
Thus, the effective Hamiltonian (\ref{7}) in $Q$-space is
non-Hermitian,
\begin{equation}
{\cal H}_{Q}=\bar{H}-\,\frac{i}{2}\,W,          \label{10}
\end{equation}
where the total Hermitian part $\bar{H}\equiv\bar{H}_{QQ}$ is a
symmetric real matrix that includes, apart from the original
Hamiltonian of $Q$-space, $H_{QQ}$, the principal value contribution
of the $QP$-coupling and in realistic applications is given by the
usual shell model, while the second part of (\ref{10}) is
anti-Hermitian.

The observable cross section of the reaction $a\rightarrow b$ is
determined by the square of the scattering amplitude
\begin{equation}
T^{ba}(E)=\sum_{qq'}\langle a|H_{PQ}|q\rangle\,\left(\frac{1}
{E^{(+)} -{\cal H}_{Q}}\right)_{qq'}\,\langle q'|H_{QP}|b\rangle.
                                                      \label{11}
\end{equation}
Here the full propagator
\begin{equation}
{\cal G}(E)=\frac{1}{E^{(+)} -{\cal H}_{Q}}            \label{12}
\end{equation}
includes all interactions between the subspaces that guarantees the
unitarity of the scattering matrix, $S=1-iT$. The eigenvalues of
${\cal H}$, ${\cal E}=E-(i/2)\Gamma$, are complex poles of the
scattering matrix corresponding to the resonances in the cross
sections. Here, however, one has to have in mind that the resulting
picture in general can be complicated due to the interference of the
resonances and dependence of the effective Hamiltonian on running
energy $E$.

\subsection{Single channel case}

To demonstrate in a simple way the effect of the anti-Hermitian term
let us assume that only one channel is open. Then the matrix $W$,
eq. (\ref{9}), has a completely separable form,
\begin{equation}
\langle q|W|q'\rangle=2\pi A_{q}^{c}A_{q'}^{c\ast},   \label{13}
\end{equation}
where the amplitudes of the continuum coupling are
\begin{equation}
A^c_{q}=\langle q|H_{QP}|c\rangle.                   \label{14}
\end{equation}

The rank of the factorized matrix $W$, eq. (\ref{13}), equals 1, so
that all eigenvalues of $W$ are zero, except one that has the value
equal to the trace $\Gamma_{0}$ of this matrix,
\begin{equation}
\Gamma_{0}=\sum_{q}\langle q|W|q\rangle=2\pi\sum_{q}
|A^{c}_{q}|^{2}.                                      \label{15}
\end{equation}
Let us assume that the Hermitian part $\bar{H}$ of the effective
Hamiltonian (\ref{10}) is diagonalized, its eigenstates are
$|g\rangle$ and the eigenvalues $E_{g}$ are degenerate. Making the
additional diagonalization of $W$, we single out one superposition
of the states $|g\rangle$, namely the decaying state with the width
(\ref{15}). while the orthogonal states are still degenerate and
stable corresponding to the zero eigenvalues of $W$. The special
unstable superposition of the eigenstates is often referred to as
the {\sl super-radiant} (SR), in analogy to the Dicke coherent state
\cite{dicke54,VZWNMP04} of a set of two-level atoms coupled through
the common radiation field. Here the coherence is generated by the
common decay channel. The stable states are {\sl trapped} and
decoupled from the continuum.

This result was obtained under strict assumptions. However, it is
quite robust. In the eigenbasis $|g\rangle$ of the real part, the
anti-Hermitian part is still factorized, the amplitudes $A^{c}_{q}$
are transformed to new values
\begin{equation}
B^{c}_{g}=\langle g|H_{QP}|c\rangle                \label{16}
\end{equation}
but the trace (\ref{15}) is invariant. If the eigenstates
$|g\rangle$ are not degenerate but the typical spacings $D$ in their
spectrum are small, $D\ll\langle \gamma_{g}\rangle$, compared to the
typical values of the unperturbed widths
\begin{equation}
\gamma_{g}^{c}=2\pi|B^{c}_{g}|^{2},              \label{17}
\end{equation}
the qualitative picture is essentially the same \cite{SZAP92}. Among
the eigenstates of the total non-Hermitian Hamiltonian, there is
still one broad state and the rest of the states are very narrow. In
fact, one can start with the diagonalization of $W$ \cite{SZAP92}.
In this {\sl doorway basis} one state concentrates the whole width
while the diagonalization of $\bar{H}$ in the orthogonal subspace
defines the trapped states. The remaining Hermitian coupling
transfers small widths to the trapped states.

\subsection{General case}

The phenomenon of super-radiance survives in a general situation of
$N$ intrinsic states and $N_{c}$ open channels provided $N_{c}\ll
N$, if the mean level spacing $D$ of internal states and their
characteristic decay widths $\gamma^{c}$ satisfy the same condition
as earlier,
\begin{equation}
\kappa^{c}=\frac{\gamma^{c}}{D}> 1.               \label{18}
\end{equation}
In the opposite limit, $\kappa^{c} \ll 1$, the anti-Hermitian part
$W$ is a weak perturbation that provides small widths to intrinsic
states $|g\rangle$ converting them into isolated narrow resonances.
In this limit the quantities (\ref{17}) are partial widths for a
certain decay channel, and the total widths of resonances are given
by
\begin{equation}
\Gamma_{g}=\sum_{c}\gamma_{g}^{c}.          \label{19}
\end{equation}

As the parameters $\kappa^{c}$ increase and reach the order of one,
the influence of corresponding channels ceases to be a weak
perturbation and the term $W$ starts to dominate the dynamics. The
resonances go over to the overlapping regime and their interaction
through the common continuum channels leads to the restructuring of
the complex energy spectrum, similarly to the formation of the Dicke
coherent state. The picture in this limit simplifies with the
primary diagonalization of $W$. Since the rank of this factorized
matrix is equal to $N_{c}$, it has only $N_{c}$ non-zero positive
eigenvalues $w_{s},\; s=1,...,N_{c}$. As it is easy to derive from
eq. (\ref{13}), these nontrivial eigenvalues are at the same time
the eigenvalues of the Hermitian $N_{c}\times N_{c}$ matrix in the
channel space,
\begin{equation}
X^{ba}=\sum_{q}A^{b\ast}_{q}A^{a}_{q}.       \label{20}
\end{equation}

The intrinsic space $\{Q\}$ is now divided into the $SR$ subspace
$|s\rangle$ of dimension $N_{c}$ and the subspace of trapped states
$|t\rangle$ of dimension $N-N_{c}$ that can couple to the continuum
only through the Hermitian interaction $H_{st}$ with the states
$|s\rangle$ of the first class. In the regime of strong continuum
coupling, the interaction between the two blocks is weak
\cite{SZAP92}. A simple perturbation theory gives small widths to
the trapped states,
\begin{equation}
\Gamma_{t}=\sum_{s=1}^{N_{c}}\frac{w_{s}|H_{st}|^{2}}
{(\Delta\epsilon_{st})^{2}+w_{s}^{2}/4}.           \label{21}
\end{equation}
For example, returning again to the one-channel case
($s=1,\;t=2,..., N$) when the only non-zero eigenvalue of $W$ equals
$w_{1}=\Gamma_{0}$, eq. (\ref{15}), and the eigenvalues of $\bar{H}$
in the trapped block are $\epsilon_{t}$, we have
\begin{equation}
{\cal H}_{11}=\epsilon_{1}-\frac{i}{2}\,\Gamma_{0}, \label{22}
\end{equation}
and the characteristic equation for complex energies ${\cal E}$
takes the form
\begin{equation}
{\cal E}-\epsilon_{1}+\frac{i}{2}\,\Gamma_{0}-\sum_{t}\frac{|
H_{1t}|^{2}}{{\cal E}-\epsilon_{t}}=0.      \label{23}
\end{equation}
In the limit (\ref{18}), eq. (\ref{23}) gives for the imaginary part
of narrow states
\begin{equation}
\Gamma_{t}=\Gamma_{0}\,\frac{|H_{1t}|^{2}}
{(\epsilon_{1}-\epsilon_{t})^{2}+\Gamma_{0}^{2}/4},    \label{24}
\end{equation}
in agreement with (\ref{21}).

The segregation of the two classes of states can be interpreted as
physical separation of the time scales corresponding to direct and
compound processes. This separation of widths was noticed
\cite{kleinwachter85} in numerical calculations and explained in
terms of the effective Hamiltonian in \cite{SZNPA89}. The
segregation picture in the complex energy plane becomes spectacular
in the many-channel case \cite{izr94}. The representing set of
resonances in the complex energy plane at
$\langle\kappa^{c}\rangle\sim 1$ undergoes a sharp phase transition
from the uniform cloud around the average small width to two well
separated clusters, $(N-N_{c})$ trapped states near the real energy
axis and $N_{c}$ broad SR states.

\section{Doorways}

In order to see the emergence of intermediate structure we consider
a frequent situation when only a subset of intrinsic states $\{Q\}$
connects directly to the $\{P\}$ space of channels. The rest of
states in $\{Q\}$ will connect to $\{P\}$ only when they obtain
admixtures of these selected states of the first type. The special
states directly coupled to continuum are the {\sl doorways},
$|d\rangle$. They form the doorway subspace $\{D\}$ within $\{Q\}$,
and the corresponding projection operator will be denoted here as
${\cal D}$. The remaining states in $\{Q\}$ will be denoted as
$|\tilde{q}\rangle$.

The decomposition of the full Hamiltonian now reads
\[H=(H_{\tilde{Q}\tilde{Q}}+H_{DD}+H_{\tilde{Q}D}+H_{D\tilde{Q}})\]
\begin{equation}
+\,(H_{PP}+H_{DP}+H_{PD}).                        \label{25}
\end{equation}
Diagonalizing the operator in the upper line of (\ref{25}) would
give back the states $|g\rangle$ with the components of $|d\rangle$
mixed with $|\tilde{q}\rangle$ states. The two last items in the
second line of (\ref{25}) couple the states $|d\rangle$, and
therefore all states $|g\rangle$, to the open channels $|c\rangle$.

\subsection{A single doorway}

We start with the case when there is only one important doorway
state $|d\rangle$. The matrix elements of the effective operator $W$
in the intrinsic space are now given by
\begin{equation}
\langle q|W|q'\rangle=2\pi\sum_{c=1}^{N_{c}}\langle
q|H_{DP}|c\rangle\langle c|H_{PD}|q'\rangle.        \label{26}
\end{equation}
Under the doorway assumption,
\begin{equation}
\langle q|H_{DP}|c\rangle=\langle q|d\rangle\langle d|H_{DP}
|c\rangle,                                       \label{27}
\end{equation}
where $\langle q|d\rangle$ is the admixture of the doorway to the
state $|q\rangle$. Whence, eq. (\ref{26}) reduces to
\begin{equation}
\langle q|W|q'\rangle=2\pi\langle q|d\rangle\langle d
|q'\rangle\sum_{c}|\langle d|H_{DP}|c\rangle|^{2}.    \label{28}
\end{equation}
The matrix elements (\ref{2}) are again separable, this time
irrespective of the number $N_{c}$ of open channels. The doorway
state serves as a single {\sl filter} for coupling of the $\{Q\}$
space to the open channel space.

If the quantitative conditions for a super-radiant mechanism to work
are satisfied (see the discussion below) we again find one broad
state with a width
\begin{equation}
\Gamma_{s}=2\pi\sum_{q}|\langle q|d\rangle|^{2}\sum_{c}|\langle
d|H_{DP}|c\rangle|^{2}=2\pi\sum_{c}|\langle d|H_{DP}|c\rangle|^{2}.
                                                   \label{29}
\end{equation}
Naturally, the width $\Gamma_{s}$ is nothing but the total decay
width, $\Gamma^{\uparrow}_{d}$, of the doorway into all open
channels. The picture is essentially the same as in the case of one
channel although the mechanism is different. The rest of the states
in $\{Q\}$ remain stable in the limiting degenerate situation and
have small widths if the degeneracy is not perfect.

\subsection{When is this picture valid?}

As we discussed, the criterion of validity is that the average
spacing between the levels in $\{Q\}$ is smaller than the decay
width of such a state ``before" the SR mechanism was set at work. In
relation to that, we can consider the {\sl spreading width},
$\Gamma^{\downarrow}_{d}$, of the doorway state for the
fragmentation into compound states $|\tilde{q}\rangle$. If $N_{q}$
is the number of compound states in the interval covered by the
spreading width, their average energy spacing is
\begin{equation}
\bar{D}_{q}\approx \frac{\Gamma^{\downarrow}_{d}}{N_{q}}.
                                                   \label{30}
\end{equation}
Before the SR mechanism was turned on, the average decay width of a
typical $|q\rangle$ state was
\begin{equation}
\Gamma^{\uparrow}_{q}=2\pi\sum_{c}|\langle q|H_{QP}|c\rangle|^{2}
                                                  \label{31}
\end{equation}
that can be estimated as
\begin{equation}
\Gamma^{\uparrow}_{q}=\frac{\Gamma_{s}}{N_{q}}.      \label{32}
\end{equation}
Therefore,
\begin{equation}
\frac{\Gamma^{\uparrow}_{q}}{\bar{D}_{q}}\,\approx\,\frac{\Gamma_{s}}
{\Gamma^{\downarrow}_{d}}\,\approx\,\frac{\Gamma_{d}^{\uparrow}}
{\Gamma_{d}^{\downarrow}}.                          \label{33}
\end{equation}
We conclude that the requirement for the SR doorway mechanism to be
valid can be formulated as
\begin{equation}
\frac{\Gamma_{d}^{\uparrow}} {\Gamma_{d}^{\downarrow}}\,>1.
                                                    \label{34}
\end{equation}
This condition is often satisfied, compare with the discussion of a
{\sl ``broad pole"} in Refs. \cite{brentano82,SZ97}.

\subsection{Total width of a doorway state}

Now we can show that the total observed width of a doorway state is
in fact a simple sum of the continuum width $\Gamma^{\uparrow}_{d}$,
eq. (\ref{29}), and the spreading width $\Gamma^{\downarrow}_{d}$
used for estimates in the previous subsection.

The remaining states $|\tilde{q}\rangle$ of the $\{Q\}$ subspace
play the role of the background. We can diagonalize this subset of
$\{Q\}$ and get the states $|\nu\rangle$ as the eigenvectors of
$H_{\tilde{Q}\tilde{Q}}$ and their unperturbed energies $E_{\nu}$.
Let the matrix elements of the coupling of the states $|\nu\rangle$
with the doorway state be
\begin{equation}
\langle d|H_{D\tilde{Q}}|\nu\rangle\equiv h_{\nu}.     \label{35}
\end{equation}
Then the characteristic equation for the complex energies in the
$\{Q\}$ space is analogous to (\ref{23}),
\begin{equation}
{\cal E}-E_{d}+\,\frac{i}{2}\,\Gamma^{\uparrow}_{d}-\sum_{\nu}\,
\frac{|h_{\nu}|^{2}}{{\cal E}-E_{\nu}}=0,            \label{36}
\end{equation}
where $E_{d}$ is an unperturbed position of the doorway state given
by the matrix element $\langle d|H_{DD}|d\rangle$.

Under the condition of validity (\ref{34}), the sum in eq.
(\ref{36}) can be considered in a standard way \cite{BM89}.
Substituting this sum by an integral with a mean level density $\rho
=1/\bar{D}_{q}$, we acquire the real shift $\Delta$ of the doorway
state and its effective spreading width given by the {\sl ``golden
rule"},
\begin{equation}
\Gamma^{\downarrow}_{d}=2\pi\,\frac{\bar{|h|}^{2}}{\bar{D}_{q}}.
                                                 \label{37}
\end{equation}
The standard expression (\ref{37}) is valid as long as the spreading
width (\ref{37}) does not exceed the energy interval $\Delta E$
within which the level density $\rho$ and the coupling matrix
elements $h_{\nu}$ can be substituted by their average values
\cite{LewZ94,FBZ96}. As a result, the doorway state is observed in
the experiment that does not resolve the background structure of a
resonance,
\begin{equation}
{\cal E}_{d}=E_{d}+\Delta-\frac{i}{2}\,\Bigl(\Gamma_{d}^{\uparrow}
+\Gamma_{d}^{\downarrow}\Bigr).                 \label{38}
\end{equation}
Here $\Delta$ and $\Gamma_{d}^{\downarrow}$ can be smooth function
of energy taken in eq. (\ref{38}) at a doorway centroid. We can
mention also that if the background states have their own small
continuum widths $\gamma$ (for example, related to slow evaporation
processes), one needs to substitute in eq. (\ref{38})
$\Gamma^{\uparrow}_{d}$ by $\Gamma^{\uparrow}_{d}-\gamma$ since the
fragmentation of the doorway state depends on its real coupling to
the background rather than on a total flow outside through the
background.

\section{Examples}

\subsection{Isobaric analog state (IAS)}

The isobaric analog state (IAS), $|A\rangle$, is defined
\cite{auerbach72} as a result of action by the isospin lowering
operator onto a parent state $|\pi\rangle$ of a certain isospin $T$,
\begin{equation}
|A\rangle={\rm const}\,\cdot T_{-}|\pi\rangle.          \label{39}
\end{equation}
In a compound nucleus, the IAS is surrounded by many compound states
$|q\rangle$ of lower isospin $T_{<}=T-1$. The Coulomb interaction
violates the isospin purity fragmenting the strength of the IAS over
many states $|q\rangle$ that results in the spreading width
$\Gamma^{\downarrow}_{A}$ of the IAS. If located above thresholds,
the IAS can also decay into several continuum channels that gives
rise to the decay width $\Gamma^{\uparrow}_{A}$. In heavy nuclei,
usually
\begin{equation}
\Gamma^{\uparrow}_{A}>\Gamma^{\downarrow}_{A}.      \label{40}
\end{equation}
For example, in the lead region, $\Gamma^{\uparrow}_{A}/
\Gamma^{\downarrow}_{A}\approx 2$ \cite{auerbach72,reiter90}.

The SR mechanism is therefore relevant to this case providing a
straightforward explanation why the IAS appears as a single
resonance with the decay width given by that of $|A\rangle$,
\begin{equation}
\Gamma^{\uparrow}_{IAS}=2\pi|\langle A|H_{QP}|P\rangle|^{2}.
                                               \label{41}
\end{equation}

\subsection{Single-particle resonance}

Let us consider a single-particle state $|\phi_{{\rm s.p.}} \rangle$
that belongs to the space $\{Q\}$ but has energy embedded in the
continuum (above the decay threshold). Such a single-particle state
couples directly to the continuum channels, often through a one-body
mean field potential $V$. Due to the residual interaction, the
single-particle state will be spread over complicated compound
states. At low energy above threshold, the main coupling of the
compound states $|q\rangle$ to the continuum channels takes place
via the single-particle component $\langle \phi_{{\rm
s.p.}}|q\rangle$. Thus, the state $|\phi_{{\rm s.p.}}\rangle$ serves
as a doorway and therefore our discussion of the single doorway case
applies to this situation.

This mechanism will produce a state with a large single-particle
width given by
\begin{equation}
\Gamma_{{\rm s.p.}}=2\pi\sum_{c}|\langle\phi_{{\rm s.p.}} |V|
c\rangle|^{2}.                                  \label{42}
\end{equation}
This width should be identified with the width of single-particle
resonances known in the framework of the optical potential in
Feshbach theory \cite{feshbach5862}. The narrow neutron resonances
enveloped by a single-particle resonances represent the trapped
states. In the modern versions of the continuum shell model based on
the effective non-Hermitian Hamiltonian \cite{VZCSM05}, this
approach, in the combination with the spectroscopic amplitudes
describing the spreading into complicated configurations, allows one
to find the widths of individual resonances and the particle cross
section as a function of energy. Here the threshold energy
dependence of the width (\ref{42}) cannot be ignored
\cite{SZAP92,VZCSM05}.

\section{Several channels and intermediate structure}

Here we return to the situation when the $|q\rangle$ states are
coupled to a number of open channels in the $\{P\}$ space, a typical
situation in nuclear physics. In addition to the elastic channel,
the compound states in many cases can decay to excited states in the
daughter nuclei.

The one-body decay channels can be described either as simple hole
states in the daughter nucleus and a particle in the continuum or
{\sl collective} particle-hole (p-h) excitations of the daughter
nucleus with the nucleon in the continuum. As mentioned above, in
such situations the number of open channels is small compared to to
the number of compound states. This leads to the correlations in the
matrix elements of $W$, and as a result we expect a number of
resonances with the widths that are of intermediate size between the
compound widths and the single-particle width.

If $|h\rangle$ is a hole excitation in the daughter nucleus, and
$|\tilde{p}\rangle$ the state of a nucleon in the continuum, the
channel states can be presented as $|c\rangle=|h;\tilde{p}\rangle$.
Then the $W$ matrix takes the form
\begin{equation}
\langle q|W|q'\rangle=\sum_{h,\tilde{p}}\langle
q|V|h;\tilde{p}\rangle\langle h;\tilde{p}|V|q'\rangle. \label{43}
\end{equation}
Among the various configurations contained in the $\{Q\}$ space
there are the ones which have the particle-hole structure,
\begin{equation}
|d\rangle=|h;p\rangle,                          \label{44}
\end{equation}
with the {\sl bound} single-particle state $|p\rangle$. These states
are mixed into the rest of compound states with the amplitudes
$\langle q|h;p\rangle$. This mixing will be dominated by the
coupling of these components to the above defined channels
$|h;\tilde{p}\rangle$. Then the dominant contribution to the matrix
elements in eq. (\ref{43}) will be
\begin{equation}
\langle q|W|q'\rangle=\sum_{h,p,\tilde{p}}\langle q|h;p\rangle
\langle h;p|V|h;\tilde{p}\rangle\cdot\langle h;\tilde{p}|V|h;p
\rangle\langle h;p|q'\rangle.                 \label{45}
\end{equation}
The sum here includes also  the states with the same hole
$|h\rangle$ but different particle states $|p\rangle$.

We should stress here the difference between this case and the one
discussed earlier of a single doorway and many channels. In the
latter case, the matrix $W$ was factorized as $W_{qq'}=a_{q}
a^{\ast}_{q'}$. In the present case the $W$ matrix cannot be written
in this simple form because each channel has its own doorway(s)
$|d\rangle$, so that the form of the matrix $W$ is
\begin{equation}
W_{qq'}=\sum_{d=1}^{N_{c}}a_{q}^{d}a_{q'}^{d\ast}. \label{46}
\end{equation}
If $N_{c}\ll N_{q}$, we have correlations in $W$. This scenario of
several open channels of specific nature leads to a cross section
that exhibits resonances with widths $\Gamma_{{\rm int}}$ of
intermediate size, $\Gamma_{q}\ll \Gamma_{{\rm int}}<\Gamma_{{\rm
s.p.}}$. Such resonances were observed in nuclear reactions long ago
\cite{holbrow63,wood65} and the cross sections were said to exhibit
intermediate structure \cite{kerman63,ferrell66}.

Intermediate structure of a different type, even for a single
channel case, can arise as a signature of the presence of intrinsic
states of different degree of complexity. This is a time-independent
description of the conventional multi-step dynamics. Let us assume
that the doorway state $|d\rangle$ is coupled directly only with a
subset $|\nu\rangle$ of intrinsic states. Those can be, for example,
particle-hole states with a straight connection to the
single-particle resonance. There exist more complicated intrinsic
states $|\lambda\rangle$ coupled with the p-h states through the
matrix elements $h_{\nu\lambda}$ but decoupled from the doorway.
This leads to the appearance of a hierarchy of spreading with the
fragmentation of the first generation states $|\nu\rangle$. The
standard consideration shows that, for incoherent couplings
$h_{\nu\lambda}$, the propagator $G(E)=1/(E-H)$ has in the p-h space
matrix elements
\begin{equation}
G_{\nu\nu}(E)=\frac{1+|h_{\nu}|^{2}/\{(E-\epsilon_{\nu})[E-E_{d}+
(i/2)\Gamma^{\downarrow}_{d}]\}}{E-\epsilon_{\nu}+(i/2)
\Gamma^{\downarrow}_{\nu}}.                   \label{47}
\end{equation}
Here, along with the spreading width $\Gamma^{\downarrow}_{d}$ of
the doorway state, eq. (\ref{37}), the spreading width of p-h states
appears,
\begin{equation}
\Gamma^{\downarrow}_{\nu}=2\pi\,\frac{\langle |h_{\nu\lambda}|^{2}
\rangle} {D_{\lambda}},                                \label{48}
\end{equation}
where $D_{\lambda}$ is the mean level spacing of the states
$|\lambda\rangle$. Due to the large density $1/D_{\lambda}$, the
widths (\ref{48}) of different states $|\nu\rangle$ should, as a
rule, overlap, so that for their extraction from data one needs to
perform additional statistical analysis, for instance along the
lines of Refs. \cite{kilgus87,shevchenko04,kalmykov06}.

In some situations, several doorway states $|d\rangle$ appear
separated by energy intervals larger than their total widths. The
intrinsic states $|q\rangle$ couple to these doorways locally giving
rise to spreading of doorways. According to the doorway hypothesis,
the $|q\rangle$ states couple to the continuum only via the
admixture(s) of the local doorways. Then the matrix $W$ separates
into several disconnected quadratic blocks of dimensions
$n_{d}\times n_{d}$ along the diagonal, where $n_{d}$ is the number
of states $|q\rangle$ contained within the spreading width of the
doorway $|d\rangle$.

For each such a submatrix, eq. (\ref{46}) is valid, and the problem
is reduced to that of the single doorway. The resulting picture is a
series of resonances separated in energy with decay widths equal to
the decay widths of individual doorways $|d\rangle$.

\section{Giant resonances}

Giant resonances in nuclei (or atomic clusters) give another example
of similar physics \cite{BBB98}. Usually, the giant resonances are
discussed in terms of collectivization of p-h configurations with
identical spin-parity quantum numbers. The coherent residual
interactions form a correlated state that carries much of the
transition strength of a corresponding multipole operator. The giant
resonances are mostly located in the particle continuum decaying via
particle emission to the ground and excited states in the daughter
nucleus.

Since the p-h giant resonance $|G\rangle$ is surrounded usually by a
dense spectrum of 2p-2h and more complex configurations, the
residual strong interaction will mix $|G\rangle$ with this
background. Each of the resulting states denoted as $|b\rangle$ will
contain the admixture $\langle G|b\rangle$ of the giant resonance.
The mixed states $|b\rangle$ couple to the continuum. If we assume
that the dominant coupling is that through the admixture of the
giant state, then $|G\rangle$ serves as a doorway, and the matrix
$W$ is given by
\begin{equation}
\langle b|W|b'\rangle=\langle b|G\rangle\langle G|b'\rangle\cdot
\sum_{c=1}^{N_{c}}\langle G|V|c\rangle\langle c|V|G\rangle.
                                                   \label{49}
\end{equation}
Again, the matrix $W$ for the background states is of rank one, and
the SR will emerge with the decay width
\begin{equation}
\Gamma_{G}^{\uparrow}=2\pi\sum_{c}|\langle G|V|c\rangle|^{2}.
                                                \label{50}
\end{equation}

This is valid under the assumption that the energy spread of the
background states $|b\rangle$ is small compared to their decay
width,
\begin{equation}
\Gamma_{b}^{\uparrow}=2\pi\sum_{c}|\langle c|V|b\rangle|^{2}.
                                                \label{51}
\end{equation}
As before, this condition can be expressed as
\begin{equation}
\frac{\Gamma_{G}^{\uparrow}}{\Gamma_{G}^{\downarrow}}\,>1,
                                               \label{52}
\end{equation}
the spreading width of the giant resonance is smaller than its total
decay width. When this condition is not satisfied, so that the
spacings between the $2p-2h$ states are larger than their decay
widths $\Gamma_{b}^{\uparrow}$, the situation of a single decay peak
for the giant resonance might not hold. Still one could expect some
bunching of background states to group in energy and have spacings
within the group smaller than their decay widths
$\Gamma_{b}^{\uparrow}$. Each such group then can be treated
separately using the SR mechanism and appear as a single peak in the
decay (or excitation) curve. Then one will observe intermediate
structure resonances in the GR energy domain.

\section{Double humped potentials}

As a function of a collective coordinate, double (or even triple)
humped potentials are often encountered, especially in heavy nuclei.
For example, such structure in the dependence of collective energy
on the quadrupole deformation parameter $\beta_{2}$ corresponds to
normal-deformed and superdeformed wells. The first potential is
usually deeper, and the decay of intrinsic states $|q\rangle$ in
this potential is hampered because they have to penetrate the double
barrier when decaying to the continuum.

The states defined in the first well can couple to a few states
$|d\rangle$ in the shallower second potential that couples to the
continuum through a single outer barrier. These states $|d\rangle$
serve as doorways for the decay of states $|q\rangle$ in the first
potential well. Similar to the previous examples, the matrix $W$ can
be written in the form
\begin{equation}
\langle q|W|q'\rangle=\sum_{cdd'}\langle q|d\rangle\,\langle
d|V|c\rangle\,\langle c|V|d'\rangle\,\langle d'|q'\rangle.
                                                   \label{53}
\end{equation}
If the number of doorways $|d\rangle$ is small compared to the
number of states $|q\rangle$, one should expect, as a result of
diagonalization of the above matrix, several broader states with the
rest of the states being narrow.

An outstanding example are the fission isomers
\cite{polikanov68,weigmann68,BM89} in several heavy nuclei, such as
$^{241}$Pu. The fission cross section in the neutron induced
reaction on $^{240}$Pu shows the structures with larger widths
spaced several hundred eV apart, whereas the average spacing between
the compound resonances is of the order of eV. This phenomenon is
interpreted as a direct consequence of a double humped potential in
the fission process of a compound nucleus. The energy spacings
between the few excited (usually, collective - rotational or
vibrational) states in the second shallower potential are larger
than between the compound states in the deeper well. When the
compound states in the deeper well are in the energy vicinity of the
states in the second well, they couple forming mixed states which,
in turn, couple to the fission channel via the admixture of the
state from the second well.

In the formalism we apply in this work, the states in the second
minimum are the doorways $|d\rangle$ for the decay of a compound
state into the open channels, in this case the fission ones,
$|f\rangle$. The matrix $W$ for the compound states $|q\rangle$ in
the vicinity of the doorway $|d\rangle$ can be written as
\begin{equation}
\langle q|W|q'\rangle=\langle q|d\rangle\,\langle d|q'\rangle\,
\sum_{f}|\langle d|V|f\rangle|^{2}.               \label{54}
\end{equation}
After the diagonalization of this matrix, one will find a state with
a considerably wider decay width than that of a typical compound
state but of course smaller than a single-particle decay width. The
situation repeats itself  for each of the states in the second well.
As a result, one will observe intermediate structure in the fission
cross section.

For the gamma decay from the second (superdeformed) well to the
normal-deformed states, see \cite{sargeant05} and references
therein, the role of a similar doorway can be played by a collective
multi-phonon state from the first well that has a large vibrational
amplitude and therefore a noticeable tail stretched into the second
well.

\section{Conclusion}

We have considered few examples justifying a generic qualitative
picture of formation of doorway states and intermediate structures
in various nuclear processes. The traditional theory of intermediate
structure \cite{griffin66, feshbach80,friedman81} considers usually
statistical sequences in time of multi-step particle interactions
with possible precompound (or preequilibrium) decay to the
continuum. The present formulation emphasizes complementary aspects
of underlying physics, namely those that follow from the strict
quantum-mechanical description of complicated many-body dynamics
generated by the mean field and residual interactions. The approach
based on the effective non-Hermitian Hamiltonian rigorously predicts
the existence of different scales in the energy dependence of
observed cross sections. These scales are defined by the intrinsic
dynamics that combines collective (coherent) and chaotic
(incoherent) features. The new collectivity responsible for the
emergence of the doorways and related intermediate structure
appears, under certain physical conditions, due to the factorized
nature of corresponding terms in the effective Hamiltonian. This
separability is deeply related to the properties of unitarity of the
scattering matrix. In this way we obtain a supplementary description
of many-body dynamics.\\
\\
{\small Support from the National Science Foundation grant
PHY-0555366 and the grant from the Binational Science Foundation
USA-Israel is gratefully acknowledged.}

\end{document}